\documentclass[prb,aps,superscriptaddress,twocolumn,notitlepage,showpacs,nofootinbib]{revtex4-1}
\usepackage{latexsym}
\usepackage{amsmath, dsfont, physics}
\usepackage{amssymb}
\usepackage{graphicx}
\usepackage{caption}
\usepackage{subfigure}
\usepackage{float}
\usepackage{mathrsfs}
\usepackage{color}
\usepackage{txfonts}
\usepackage[justification=centering,
            format=plain]{caption}

\renewcommand{\raggedright}{\leftskip=0pt \rightskip=0pt plus 0cm}

\begin{document}

\title{Cavity mediated level attraction and repulsion between magnons}

\author{Jayakrishnan M. P. Nair}
\email{jayakrishnan00213@tamu.edu}
\affiliation{Institute for Quantum Science and Engineering, Texas A$\&$M University, College Station, TX 77843, USA}
\affiliation{Department of Physics and Astronomy, Texas A$\&$M University, College Station, TX 77843, USA}

\author{Debsuvra Mukhopadhyay}
\email{debsosu16@tamu.edu}
\affiliation{Institute for Quantum Science and Engineering, Texas A$\&$M University, College Station, TX 77843, USA}
\affiliation{Department of Physics and Astronomy, Texas A$\&$M University, College Station, TX 77843, USA}
%\affiliation{Texas A$\&$M University, College Station, TX 77843, USA}

\author{Girish S. Agarwal}
\email{girish.agarwal@tamu.edu}
\affiliation{Institute for Quantum Science and Engineering, Texas A$\&$M University, College Station, TX 77843, USA}
\affiliation{Department of Physics and Astronomy, Texas A$\&$M University, College Station, TX 77843, USA}
\affiliation{Department of Biological and Agricultural Engineering, Texas A$\&$M University, College Station, TX 77843, USA}
%\affiliation{Texas A$\&$M University, College Station, TX 77843, USA}

\date{\today}

\begin{abstract}
We characterize some of the distinctive hallmarks of magnon-magnon interaction mediated by the intracavity field of a microwave cavity, along with their testable ramifications. In general, we foreground two widely dissimilar parameter domains that bring forth the contrasting possibilities of level splitting and level crossing. The former is observed in the regime of strong magnon-photon couplings, particularly when the three modes bear comparable relaxation rates. This character is marked by the appearance of three distinguishable and non-converging polariton branches in the spectral response to a cavity drive. However, when the bare modes are resonant and the couplings perfectly symmetrical, one of the spectral peaks gets wiped out. This anomalous extinction of polaritonic response can be traced down to the existence of a conspicuous dark mode alongside two frequency-shifted bright modes. In an alternate parameter regime, where the magnon modes are weakly coupled to the cavity, features of level attraction unfold, subject to a large relaxation rate for the cavity mode. Concurrently, for antisymmetric detunings to the magnon modes, a transmission window springs into existence, exhibiting transparency in the limit of negligible dissipation from the magnons. The emergence of level attraction can be reconciled with a theoretical model that embodies the dynamics of the magnon-magnon subsystem when the cavity field decays rapidly into its steady state. In this limit, we identify a purely dissipative coupling between the magnon modes.
\end{abstract}
\maketitle

\section{Introduction}
The expedient control of coherent magnon-photon coupling in hybrid cavity-spintronic systems has laid bare a plethora of new avenues for magnon-based quantum information transfer \cite{PhysRevLett.111.127003,PhysRevLett.113.083603,PhysRevLett.113.156401, PhysRevLett.114.227201, zhang2015magnon, tabuchi2015coherent, chumak2015magnon, zhang2016cavity, PhysRevLett.116.223601, PhysRevLett.118.217201, zhang2017observation, PhysRevA.99.021801, PhysRevLett.121.203601, PhysRevResearch.1.023021, nair2020deterministic, PhysRevLett.120.057202, lachance2019hybrid, PhysRevB.102.104415, PhysRevB.103.224401}. The collective spin-wave excitations called magnons in magnetic systems can efficiently interface with microwave photons, thereby consolidating the strength of dispersive interaction to produce well-separated hybridized states demonstrating level repulsion and Rabi oscillations \cite{khitrova2006vacuum, PhysRevLett.53.1732}. Complementary to the phenomenon of coherent magnon-photon coupling in a cavity, there exists a paradigm of dissipative coupling wherein the magnon mode interacts with the modes of a waveguide \cite{PhysRevLett.121.137203, PhysRevB.100.214426}. In contrast to the mode anticrossing in dispersively coupled systems, dissipative coupling induces level attraction characterized by the coalescence of the polariton modes at the exceptional points (EPs) \cite{PhysRevB.99.134426, PhysRevApplied.11.054023, rao2019level, PhysRevB.100.094415, PhysRevLett.123.127202}. Considering the ubiquity of both coherent and dissipative couplings in cavity-magnonics as well as the wide disparity in the resulting physics, the competing effects of these couplings have evoked significant interest of late. Consequently, cavity-magnonics has emerged as a premier platform to translate some of the foremost ideas of non-Hermitian optics into scalable quantum technologies with novel, multitasking capabilities. By exploiting strongly coupled magnon-photon systems, a plenitude of exotic effects have been brought to the fore, some of which include single-spin detection \cite{lachance2017resolving} and manipulation of spin current \cite{PhysRevLett.114.227201, PhysRevB.94.054433}, polaritonic dark modes \cite{zhang2015magnon, PhysRevA.93.021803} and magnonic quintuplets \cite{yao2017cooperative}, observation of coherent perfect absorption and exceptional point \cite{zhang2017observation}, multistability \cite{PhysRevLett.120.057202,PhysRevB.102.104415},  squeezing \cite{PhysRevA.99.021801} and entanglement \cite{PhysRevLett.121.203601,PhysRevResearch.1.023021,nair2020deterministic}, sensing of anharmonicities \cite{PhysRevLett.126.180401}, and bidirectional microwave-to-optical transduction \cite{PhysRevB.93.174427, PhysRevB.102.064418, mukhopadhyay2021anti}.

Owing to its potential applications in topological energy transfer \cite{dembowski2001experimental, doppler2016dynamically, xu2016topological} and quantum sensing \cite{PhysRevLett.112.203901, chen2017exceptional, el2018non, ozdemir2019parity, miri2019exceptional, zhang2020breaking}, there has been a proliferation of efforts within the optics and condensed matter community to realize level attraction in hybrid non-Hermitian systems. So far, this effect has been successfully showcased in cavity-optomechanical systems \cite{PhysRevA.98.023841}, and cavity-magnonic configurations integrated with waveguides \cite{PhysRevLett.125.147202} and split-ring resonators \cite{PhysRevB.99.134426, PhysRevApplied.11.054023}. It is well known that two coherently coupled harmonic systems naturally exhibit level repulsion while dissipatively coupled systems are characterized by level attraction. Therefore, a customary technique to implement level crossing lies in mitigating the effect of dispersive interaction relative to its dissipative counterpart. Both dispersive and dissipative couplings naturally occur in qubits interacting with electromagnetic vacuum or with the modes of a bad cavity \cite{agarwal1974quantum}. 

In the context of magnonics, magnon-photon level repulsion and level attraction have both been experimentally realized  \cite{PhysRevLett.121.137203, PhysRevB.100.214426, PhysRevB.99.134426, PhysRevApplied.11.054023, rao2019level, PhysRevB.100.094415, PhysRevLett.123.127202}. In this work, we explore the distinctive qualities of magnon-magnon hybridization in a system of two non-interacting magnetic systems coherently coupled to a microwave cavity. We show that the mediating effect of the intracavity field brings to bear the dual possibilities of effecting both level anticrossing and level crossing in two distinct parameter regimes. Much to our anticipation, strong magnon-photon couplings result in well-separated polaritonic states, or normal modes, characterizing level repulsion. Curiously, however, we observe that the central peak is washed out when the frequencies of the bare modes are matched to resonance and the magnons are symmetrically coupled to the cavity mode. As the symmetry is broken and the magnons are antisymmetrically detuned relative to the cavity mode, the dark polariton gets reexcited on account of energy transfer from the bright modes. The contrary feature of level attraction transpires when the cavity leakage rate strongly dominates the dynamics, i.e., the magnon-photon couplings as well as the magnonic damping rates are much weaker than the cavity relaxation rate. In the same parameter regime, on examining the spectral properties of the hybrid system, we unveil a narrow transmission window, with the possibility of transparent behavior in the limit of negligible magnon dissipation. Both these features, i.e., the inception of level crossing and the appearance of transparency, are borne out by a purely dissipative form of interaction that ensues between the magnons when the cavity field is adiabatically eliminated. In fact, the adiabatic limit reproduces the characteristic dynamics of two spatially separated systems coupled through an interceding reservoir, thereby unveiling an equivalence with waveguide QED. It should be noted that qubits get coherently coupled with each other in a dispersive cavity \cite{PhysRevA.56.2249}. In a cavity-magnonic setup, such a coupling has been experimentally realized in the dispersive limit, demonstrating level repulsion between the magnons \cite{PhysRevA.93.021803}. In contrast, the level repulsion discussed in our paper makes no reference to the dispersive regime, and the magnons could well be on resonance with cavity mode.

This paper is organized as follows. In section \ref{sec1}, we spell out the theoretical model describing a pair of magnetic systems coupled to a microwave cavity, and highlight the two disparate regimes in which level splitting and level crossing could be observed. Following this, we study the induced magnetization in response to a cavity drive in section \ref{sec2}, to reaffirm the contrasting spectral signatures to both these behaviors. In section \ref{sec3}, we address the emergence of a dark polaritonic state and furnish the physical mechanism governing the excitation of this mode. Finally, in section \ref{sec4}, we illustrate the generation of a transmission window as a consequence of dissipative coupling between the two magnons.
\begin{figure}
 \captionsetup{justification=raggedright,singlelinecheck=false}
 \centering
   \includegraphics[scale=0.48]{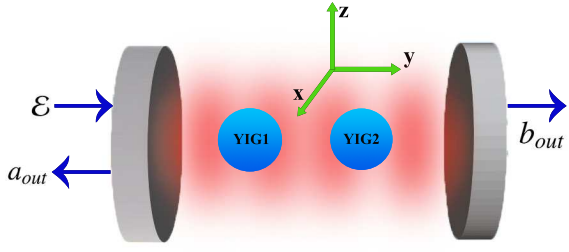}
\caption{Schematic of two YIGs coherently coupled to a single-mode microwave cavity. The static magnetic field exciting the Kittel mode in both the YIGs are aligned along the $z$-axis. The intracavity field mode is chosen to be polarized along the $y$-axis, with the corresponding magnetic field directed along the $x$-axis.}
\label{sch}
\end{figure}

\section{Theoretical Model}\label{sec1}
Our model comprises a hybrid cavity-magnonic system, in which two macroscopic yttrium iron garnet (YIG) samples are placed inside a microwave cavity, as illustrated in Fig. 1. The YIG is a high-spin-density magnetic material, and the low lying excitations of the collective angular momentum of these spins in such magnetically ordered systems gives rise to quasi-particles, namely magnons. In our model, we consider a spatially uniform Kittel mode of either YIG as being dispersively coupled to the intracavity photons of a neighboring frequency. Therefore, in essence, the magnetization associated with the YIG sphere can simply be written as the magnetization of a ferromagnet with a single large spin $\textbf{S}$, i.e., $\textbf{M}=\gamma_e\textbf{S}/V$, where $\gamma_e=e/{m_e c}$ is the gyromagnetic ratio for electron spin and $\textbf{S}$ denotes the collective spin operator, and $V$ the volume of the YIG sphere. The Hamiltonian of the hybrid magnon-cavity system is then provided by
\begin{equation}
\begin{split}
H/\hbar = -\gamma_e \sum_{i=1}^2 & B_{0} S_{i,z} + \omega_a a^{\dagger}a - \gamma_e\sum_{i=1}^2 S_{i,x}B_{i,x},
\end{split}
\label{h}
\end{equation}
where $\textbf{B}_0$ is the applied bias magnetic field along the $z$ direction, $S_{i,x}$ $(S_{i,z})$ denotes the $x$- ($z$-) component of the collective spin $\textbf{S}$ of the $i^{th}$ YIG, $\omega_a$ is the cavity resonance frequency, and $a$ $(a^{\dagger})$ represents the annihilation (creation) operator of the cavity field. The magnetic field of the cavity field ($B_{n,x}$) is assumed to be along the $x$ direction. A typical YIG sample having diameter $d=1$ mm and an approximate spin density $\rho=4.22\times 10^{27}$ m$^{-3}$ leads to the total number of spins $N\approx 10^{18}$. In the limit of a high spin density, we invoke the Holstein-Primakoff transformation to recast the raising and lowering operators as $S_i^+=\sqrt{2S_i}m_i,\ S_i^-=\sqrt{2S_i}m_i^{\dagger}$, with $i=1,2$ labeling the two YIGs) and $m_i$ ($m_i^{\dagger}$) operators satisfying the Bosonic algebra. With such simplifications, we can now tailor the Hamiltonian in Eq. (\ref{h}) into its corresponding operator representation as
\begin{equation}
H/\hbar = \omega_a a^{\dagger}a + \sum_{i=1}^2\Big[\omega_i m_i^{\dagger}m_i + g_i( m_i^{\dagger}a+m_ia^{\dagger})\Big]\label{Heff}
\end{equation}
where the frequency of Kittel mode $\omega_i = \gamma_e B_{0}$ with $\gamma_e/2\pi=28$ $\text{GHz}/\text{T}$. The parameter $g_i=\frac{\sqrt{5}}{2}\gamma_e\sqrt{N}B_{\text{vac}}$ quantifies the coherent magnon-cavity coupling, with $B_{\text{vac}}=\sqrt{2\pi\hbar\omega_c/\text{V}}$ denoting the magnetic field of vacuum. 
 \begin{figure}
 \captionsetup{justification=raggedright,singlelinecheck=false}
 \centering
   \includegraphics[scale=0.47]{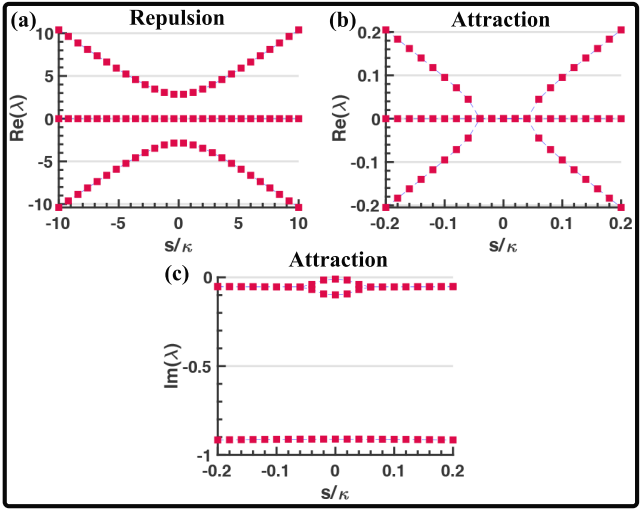}
\caption{Level anticrossing and level crossing in two different regimes of cavity QED; (a) Eigenfrequencies of the cavity-magnon system for $g=2$ and $\gamma_{i}=1$ exhibiting level splitting; (b) Eigenfrequencies and (c) Linewidths in the case of $g=0.2$ and $\gamma_{i}=0.01$, with the features of level crossing. All the parameters are chosen in units of $\kappa$.}
\label{sch}
\end{figure}
%Note that there is no external drive applied onto the system and therefore, the high spin density limit $S\gg\langle m^{\dagger}m\rangle$ is easily satisfied

In the frame rotating at frequency $\omega_a$, the effective non-Hermitian Hamiltonian describing the time evolution of the column matrix $X$, where $X^{T}=$ [$\expval{a}$ \hspace{1mm} $\expval{m_1}$ \hspace{1mm} $\expval{m_2}$] is provided by 
\begin{align}
\mathcal{H}=\begin{pmatrix}
-i\kappa&g_1&g_2\\
g_1&s-i\gamma_1&0\\
g_2&0&-s-i\gamma_2
\end{pmatrix},
\end{align}
where $s=(\omega_{m_1}-\omega_{m_2})/2$, and we have set $\omega_a=(\omega_{m_1}+\omega_{m_2})/2$. The parameters $\kappa$ and $\gamma_i$ denote the rates of dissipation appearing in the master equation. In what follows, we suppress the notation $\expval{.}$ in the mode amplitudes, as we are interested in a semiclassical description of the system. The three complex eigenvalues of $\mathcal{H}$, resulting from the characteristic polynomial equation 
\begin{align}
(\lambda+i\kappa)(\lambda+s+i\gamma_1)(\lambda-s+i\gamma_2)-g_1^2(\lambda+s+i\gamma_2)\notag\\
-g_2^2(\lambda-s+i\gamma_1)=0,
\end{align} 
identify the normal modes of the hybridized system, also referred to as the polaritons. Here, we present two distinct parametric regimes: (i) where $g\gg \kappa,\gamma_i$ with the features of level repulsion and (ii), the weak coupling regime $g,\gamma_i\ll\kappa$ where the eigenvalues relevant to the system demonstrates level attraction. Before delving into the two cases, we recall that a coherently coupled system typically demonstrates Rabi splitting in the strong-coupling regime. That is to say, coherent coupling opens up a gap between the normal mode frequencies of the system, and the eigenfrequencies bend away from each other. In contrast, dissipatively coupled systems exhibit level-attraction, which is characterized by smoothly converging eigenfrequencies leading to mode coalescence over a finite parameter domain. The points of convergence, known as EPs, are responsible for a variety of exotic phenomena owing to the square-root singularity of the modes at these points. We demonstrate in Fig. 2, the solutions of the characteristic equation (4) in the two above mentioned parameter regimes where level repulsion and level attraction come to light. Specifically, in case (ii), when the cavity is weakly coupled to the magnon modes and the photons decay over a much shorter time scale than the magnons, we can observe signatures of level crossing between the magnon-like polaritons. The contrasting features in Fig. 2(a, b) can be analytically understood by inspecting the normal modes in these special regimes, as spelled out below.

For $\kappa=\gamma_1=\gamma_2$ and $g_1=g_2=g$, the three eigenvalues of $\mathcal{H}$ are given by $\lambda_0=-i\kappa$, $\lambda_{\pm}=-i\kappa \pm \sqrt{s^2+2g^2}$. The imaginary parts of the eigenvalues, which quantify the corresponding linewidths, are all identical and insensitive to the variations in $s$. The real parts of these eigenvalues are plotted against $s$ in Fig. 2(a), where the minimum frequency gap between $\lambda_+$ and $\lambda_-$, equaling $2\sqrt{2}g$, is a manifestation of level repulsion in the system. Note that the minimum separation must be larger than the linewidth $\kappa$ to observe level repulsion. However, in the limit $\kappa\gg\gamma_i$ and $\kappa\gg g_i$, level crossing is introduced over a small band of frequency detunings $s$, as portrayed in Fig. 2(b, c). We have assumed that $\gamma_1=\gamma_2=\gamma$ and $g_1=g_2=g$. It is, therefore, imperative to design a cavity with low quality factor to observe this effect. Neglecting the effect of $\gamma$, which is weak in reference to $\kappa$, leads to the characteristic equation
\begin{align}
(\lambda+i\kappa)(\lambda^2-s^2)-2g^2\lambda = 0.
\end{align}
Since the eigenvalues in the absence of $g$ equal $\lambda_0=-i\kappa$, and $\lambda_{\pm}=\pm s$, it can be argued that the perturbative correction to each of them for sufficiently small $g$ would go as $\mathcal{O}(g^2)$. This consideration pins down the approximate form of $\lambda_0$ to be $-i\kappa[1-2g^2/(s^2+\kappa^2)]$. Next, in view of the fact that both $s$ and $g^2$ are small compared to $\kappa$, we approximate $\lambda_{\pm}+i\kappa \approx i\kappa$, which permits the reduction of the cubic equation into a quadratic one $\lambda_{\pm}^2+2\Gamma\lambda_{\pm} i-s^2\approx 0$, where $\Gamma=g^2/\kappa$. This yields the remaining eigenvalues
\begin{align}
\lambda_{\pm}=-i\Gamma\pm i\sqrt{\Gamma^2-s^2}.
\end{align}
The forms of $\lambda_{\pm}$ make apparent the level crossing between them, with EPs located at $s=\Gamma$. The plots in Fig. 2(b, c) are found to be in remarkable agreement with the expressions obtained above. 

In the following section, we extract spectroscopic information about the system by applying an external probe, which renders an empirical tool to observe level crossing and level anticrossing. 
 
 \begin{figure}
 \captionsetup{justification=raggedright,singlelinecheck=false}
 \centering
   \includegraphics[scale=0.45]{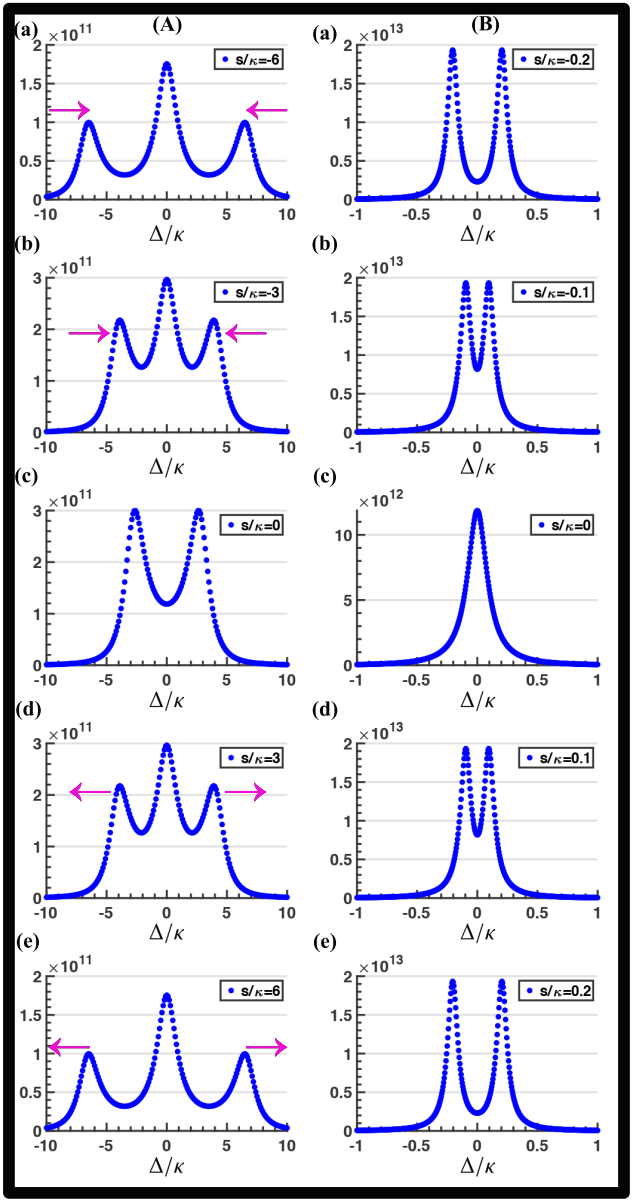}
\caption{(Panel A; a-e) Sumtotal of the spincurrent response $\abs{m_1}^2+\abs{m_2}^2$ plotted as functions of $\Delta_a$ in the case of strong magnon-photon coupling $g=2$ and equal relaxation rates $\gamma_{i}=1$ (similar to Fig. 2(a)). Note the conspicuous extinction of the central resonance in (A; c). (Panel B; a-e) Profiles for the case of $g=0.2$, $\gamma_{i}=0.01$ (similar to Fig. 2(b,c)). Only two peaks appear in this scenario. Drive power $P=1$ mW.}
\label{sch}
\end{figure}
\section{Level repulsion and attraction between magnons}\label{sec2}
In the presence of an external electromagnetic field at frequency $\omega_d$ driving the cavity, the Hamiltonian in Eq. (2) is superseded by
\begin{equation}
\begin{split}
H/\hbar = \omega_a a^{\dagger}a + \sum_{i=1}^2\Big[\omega_i m_i^{\dagger}m_i + g_i( m_i^{\dagger}a+m_ia^{\dagger})\Big]\\+i\sqrt{\kappa}\mathcal{E} [a^{\dagger}e^{-i\omega_d t}-ae^{i\omega_d t}]\label{Heff},
\end{split}
\end{equation}
with $\mathcal{E}=\sqrt{\frac{P}{\hbar \omega_d}}$ being the strength of external drive and $P$ the driving power. By switching to the rotating frame of the drive, we deduce the mean-field equations for the cavity-magnon dynamics to be
\begin{align}
\dot{X}=-i(\mathcal{H}-\Delta)X+F,
\end{align}
where we have separated out the cavity detuning $\Delta=\omega_d-\omega_a$, and $F^{T}=$$\sqrt{\kappa}$[$\mathcal{E} \hspace{1mm} 0\hspace{1mm} 0$]. The solution to the steady state can, therefore, be expressed as
\begin{align}
X(\Delta)=-i(\mathcal{H}-\Delta)^{-1}F,
\end{align}

With the poles of this expression being accorded by the eigenvalues of $\mathcal{H}$, key information regarding the polaritonic properties would be encoded in the empirically observable response, such as the steady-state spincurrents $\abs{m_1}^2$ and $\abs{m_2}^2$. In this section, we provide a numerical solution to the spectral features by scanning the probe frequency. 

In Fig. 3, on either of the panels (A) and (B), we plot the sum of the spincurrents from the individual magnon modes, i.e., $\abs{m_1}^2+\abs{m_2}^2$ as a function of $\Delta$ for five different values around $s=0$. The left panel (A) refers to the barely strong-coupling regime, i.e., $g_i>\kappa, \gamma_i$, and $\kappa=\gamma_i$. The total spincurrent, for $s\neq0$, exhibits three distinct peaks corresponding to the resonant frequencies of the system. The central peak at $\Delta=0$ is flanked on either side by two symmetrical polaritonic peaks. The disappearance of the central peak at $s=0$ is symbolic of a dark polariton \cite{yuen2019polariton}. The underlying physics of this dark mode would be discussed at length in the subsequent section. From the figures (a) to (e) in the panel (A), it is evident that the two magnon-like polaritons move closer as $s\rightarrow0$, leaving a minimum frequency gap between the two levels at $s=0$. Since the levels never intersect, it is referred to as mode anticrossing. However, this picture incurs a dramatic inversion as we transition into the domain of weaker couplings and higher relaxation rates of the cavity photons. Figure 3, panel (B), displays the sumtotal of the spincurrents as a function of $\Delta$ in the limit $\kappa\gg\gamma_i$ and $\kappa\gg g_i$, for five different values of $s$. In this scenario, around $s=0$, the magnon-like polaritons coalesce together into a single peak. Note that the cavity-like central resonance remains strongly quenched on account of its large linewidth. The onset of level crossing points to a hidden dissipative interaction between the magnon modes. In Appendix A, we establish parity with these numerical results by uncovering this dissipative magnon-magnon coupling through the adiabatic elimination of the cavity mode.

\section{Excitation of dark polariton via symmetry-breaking}\label{sec3}

As evidenced by the numerical illustration in Fig. 3, panel (A), when the magnon-photon coupling is adequately strong, the tripartite cavity-magnon system is endowed with three distinct spectral peaks characterizing the polaritonic modes of the hybrid composite. The two peripheral peaks are pushed inward as $s$ becomes smaller, settling for a minimum separation at resonance. This limiting behavior is marked simultaneously by the complete extinction of the central peak, laying bare a characteristic dark mode when the conditions conform to perfect symmetry, i.e., when all three modes have identical detunings, and the magnons are symmetrically coupled to the intracavity field. We can delineate a physical argument supporting this observation by simply inspecting the normal modes of the undamped system. In the rotating frame of the cavity field, the Hamiltonian is given by
\begin{align}
H=g[a^{\dagger}(m_1+m_2)+h.c].
\end{align}
For this Hermitian system, the polaritonic modes turn out to be $A=\dfrac{m_1-m_2}{\sqrt{2}}$, $B=\dfrac{1}{\sqrt{2}}\bigg[a+\dfrac{m_1+m_2}{\sqrt{2}}\bigg]$, and $C=\dfrac{1}{\sqrt{2}}\bigg[a-\dfrac{m_1+m_2}{\sqrt{2}}\bigg]$, corresponding respectively to the eigenfrequencies $\lambda_0=0$, $\lambda_+=\sqrt{2}g$, and $\lambda_-=-\sqrt{2}g$, which recasts the Hamiltonian into the form
\begin{align}
H=\sqrt{2}g[B^{\dagger}B-C^{\dagger}C].
\end{align}
Thus, $B$ and $C$ appear as two counteroscillating normal modes of the symmetric system, with the frequency-splitting determined by the coupling strength $g$. Note, however, the conspicuous absence of the mode $A$ in Eq. (11). This mode, therefore, represents a dark polariton, whereas $B$ and $C$ identify the bright polaritons with detectable spectral signatures. We note that dark polaritons have been extensively discussed in the context of molecules interacting with cavities \cite{yuen2019polariton}. This is what is manifested in Fig. 3 (A; c). The picture obviously changes if we begin to steer away from the symmetry, which are exemplified by the other subfigures in the same panel. For non-zero value of $s$, the Hamiltonian transforms into $s(m_1^{\dagger}m_1-m_2^{\dagger}m_2)+g[a^{\dagger}(m_1+m_2)+h.c]$. This can be expressed in the original polaritonic representation as
\begin{align}
H=\sqrt{2}g[B^{\dagger}B-C^{\dagger}C]+\frac{s}{\sqrt{2}}(A^{\dagger}B+AB^{\dagger})-\frac{s}{\sqrt{2}}(A^{\dagger}C+AC^{\dagger}).
\end{align}
The additional terms in the new Hamiltonian, originating from the frequency discord between the magnons, are clearly indicative of energy transitions from $B$ and $C$ to $A$ and vice versa. These new channels for energy exchange foreshadow how a mismatch in the magnon detunings would be instrumental to the illumination of the non-radiative polariton. The stimulating effect of symmetry-breaking can be experimentally observed via the spectroscopic properties of the system when it is irradiated by a probe field. \\

To analytically clarify the polaritonic suppression at resonance, let us examine the spectral signature of a real dissipative system. To that end, we evaluate the expressions for the steady-state response functions 
\begin{align}
m_j=-i\sqrt{\kappa}\mathcal{E}[(\mathcal{H}-\Delta)^{-1}]_{j1},
\end{align}
where $\mathcal{H}$ is the effective Hamiltonian for the $3\cross 3$ system defined in Eq. (3), and $j=1\text{ or } 2$ labels the magnon. Substituting the expression for $\mathcal{H}$ from Eq. (3), we obtain, 
\begin{align}
m_1=\frac{-i\sqrt{\kappa}\mathcal{E}g_1(s+\Delta+i\gamma_1)}{D}\\
m_2=\frac{i\sqrt{\kappa}\mathcal{E}g_2(s-\Delta-i\gamma_1)}{D},
\end{align}
 where $D=\det(\mathcal{H}-\Delta)$ given by
 \begin{align*}
 \begin{split}
 D=(\Delta+i\kappa)(s-\Delta-i\gamma_1)(s+\Delta+i\gamma_2)+g_1^2(s+\Delta+i\gamma_2)\\-g_2^2(s-\Delta-i\gamma_1).
 \end{split}
 \end{align*} In the event of symmetrical detunings ($s=0$) and equal damping rates ($\kappa=\gamma_1=\gamma_2$), the eigenvalues go as $\lambda_0=-i\kappa$, $\lambda_{\pm}=\pm \sqrt{2}g-i\kappa$. The common expression for the spincurrents then reduces to
\begin{align}
m_j=\frac{i\sqrt{\kappa}g\mathcal{E}}{(\Delta+\sqrt{2}g+i\kappa)(\Delta-\sqrt{2}g+i\kappa)},
\end{align}
for $j=1, 2$. Clearly, the contribution from the pole at $\lambda_0$ remains suppressed when the modes are on resonance and the magnon-photon couplings are perfectly symmetrical. Given the equality of the two spincurrents, it is no coincidence that the normal mode $A=\dfrac{m_1-m_2}{\sqrt{2}}$ remains unpopulated in the steady state. On the other hand, if a small asymmetry is introduced in the frequencies by letting $s$ become non-zero, the eigenvalues get modified into $\lambda_0=-i\kappa$ and $\lambda_{\pm}=\pm\sqrt{s^2+2g^2}-i\kappa$. In this case, the pole at $\lambda_0$ leaves a detectable signature as underscored by the corresponding spincurrents
\begin{align}
m_1=\frac{i\sqrt{\kappa}g\mathcal{E}(\Delta+s+i\kappa)}{(\Delta+i\kappa)(\Delta+\sqrt{s^2+2g^2}+i\kappa)(\Delta-\sqrt{s^2+2g^2}+i\kappa)},\notag\\
m_2=\frac{i\sqrt{\kappa}g\mathcal{E}(\Delta-s+i\kappa)}{(\Delta+i\kappa)(\Delta+\sqrt{s^2+2g^2}+i\kappa)(\Delta-\sqrt{s^2+2g^2}+i\kappa)},
\end{align}
which also remain unequal insofar as $s\neq 0$. This explains the emergence of the central peak in Fig. 3 as a signature of population transfer to the mode $A$. The corresponding peak height is provided by the expression
\begin{align}
\abs{m_1}^2+\abs{m_2}^2=\frac{2g^2\mathcal{E}^2(s^2+\kappa^2)}{\kappa(s^2+2g^2+\kappa^2)}.
\end{align}
Clearly, the height reduces to a minimum at $s=0$.

\section{Dissipative-Coupling-Induced Transparency}\label{sec4}
Having explored some of the distinctive regimes of cavity QED, which uncovered the contrasting features of level splitting and level crossing, it is worthwhile to study the transmission characteristics of the cavity-magnon system. For this purpose, we restrict our analysis to the regime where $\kappa\gg\gamma_i$ and $\kappa\gg g_i$, and shed light on the possibility of effecting transparent behavior. Note that this is precisely the parameter domain that elicits level attraction. 

 \begin{figure}
 \captionsetup{justification=raggedright,singlelinecheck=false}
 \centering
   \includegraphics[scale=0.47]{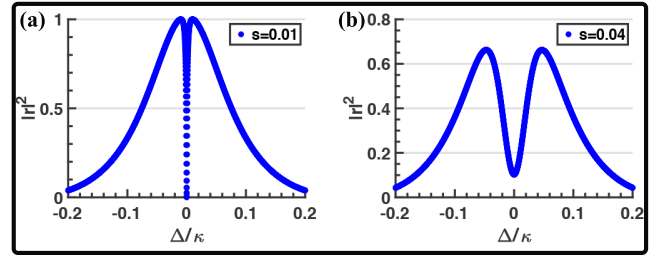}
\caption{Refection spectra for (a) $s=0.01$ with $\gamma_i=0$, and (b) $s=0.04$ with $\gamma_i=0.01$. Other parameters are $g=0.2$ and $\kappa=1$, which yields $\Gamma=g^2/\kappa=0.04$.}
\label{sch}
\end{figure}

%In the presence of an external drive on the cavity at frequency $\omega_d$, the Hamiltonian in Eq. (2) is supplemented by an additional contribution from the external field, which goes as
%\begin{equation}
%H_{int}/\hbar = i\mathcal{E} [a^{\dagger}e^{-i\omega_d t}-ae^{i\omega_d t}]\label{Heff},
%\end{equation}
%where $\mathcal{E}=\sqrt{\frac{D_p}{\hbar\omega_d}}$ represents the strength of external drive with power $D_p$. In this case, the matrix $F$ in Eq. () would be obtained as $F^{T}=\mathcal{E}$[$\sqrt{\kappa}\hspace{1mm} 0\hspace{1mm}0$]. 

For the cavity-driven system described by Eq. (7), we can obtain the the reflected and transmitted output fields from the input-output formalism, with the relevant input-output relations being
\begin{align}
b_{out}&=\sqrt{\kappa}a,\notag\\
a_{out}+\mathcal{E}&=\sqrt{\kappa}a,
\end{align}
where $a=-i\sqrt{\kappa}[(\mathcal{H}-\Delta)^{-1}]_{11}\mathcal{E}$ is the steady-state intracavity field obtained by setting $\dot{X}=0$ in Eq. (8). Substituting this into the aforementioned relations, we extract the transmission and reflection coefficents to be 
\begin{align}
t&=\frac{i\kappa(\Delta-s+i\gamma)(\Delta+s+i\gamma)}{(\Delta+i\kappa)(\Delta-s+i\gamma)(\Delta+s+i\gamma)-2g^2(\Delta+i\gamma)},\notag\\
r&=-1+\frac{i\kappa(\Delta-s+i\gamma)(\Delta+s+i\gamma)}{(\Delta+i\kappa)(\Delta-s+i\gamma)(\Delta+s+i\gamma)-2g^2(\Delta+i\gamma)}.
\end{align}
In Figs. 4(a, b), we sketch the reflection profiles for a couple of values of $s$, vindicating the existence of a transmission window. In the limit $\gamma/\kappa\rightarrow 0$, perfect transparency is observed at $\Delta=0$, as demonstrated in Fig. 4(a). Similarly, Fig. 4(b) depicts the reflection spectra at the exceptional point $s=\Gamma=0.04$ for $\gamma_i=0.01$, exhibiting a transmission window. The transparency owes its origin to a dissipative form of magnon-magnon coupling surfacing in this limit of negligible magnon dissipation. The details on this origin are elaborated in Appendix A. 
At $\Delta=0$, the expressions in Eq. (20) reduce to 
\begin{align}
t&=\frac{s^2+\gamma^2}{s^2+\gamma^2+2\gamma\Gamma},\notag\\
r&= -\frac{2\gamma\Gamma}{s^2+\gamma^2+2\gamma\Gamma},
\end{align}
where $\Gamma=g^2/\kappa$. As long as $\gamma\Gamma\ll s^2$, the approximate behavior would be given by $t\approx 1-\mathcal{O}(2\gamma\Gamma/s^2)$ and $r\approx\mathcal{O}(2\gamma\Gamma/s^2)$. 

As $s$ becomes smaller, in the limit $\gamma/\kappa\approx 0$, the EIT-like window gets progressively narrower leading to a sharpening of the central dip. This can be understood from the behavior of the eigenvalues derived earlier in Sec. II. Through pertinent approximations, we had obtained them to be $\lambda_0=-i\kappa[1-2g^2/(s^2+\kappa^2)]$ and $\lambda_{\pm}=-i\Gamma\pm i\sqrt{\Gamma^2-s^2}$. While $\lambda_0$ had rather pedestrian features, the other two normal modes $\lambda_{\pm}\approx\lambda_{\pm}=-i\Gamma\pm i\sqrt{\Gamma^2-s^2}$ were the source of level attraction. In the same vein, the unfolding of the sharp, precipitous dip in the reflection lineshapes can be traced down to the linewidths of the modes $\lambda_{\pm}$. Apropos of our parameter choices, when $s/\Gamma$ is tiny compared to $1$, as is the case in Fig. 4(a), they can be further approximated as
\begin{align}
\lambda_{+}&\approx -\frac{is^2}{2\Gamma},\notag\\
\lambda_{-}&\approx -2i\Gamma.
\end{align}
While both of them have zero real parts, $\Im(\lambda_{+})$ pales in comparison to $\Im(\lambda_{-})$. In fact, the pole at $\lambda_{+}$ has a vanishingly small linewidth, which explains the sharp and narrow trough in the reflection spectrum. The effect of $\lambda_-$, on the other hand, is embodied in the Lorentzian envelope circumscribing both the peaks flanking the central minimum.

\section{Conclusions}

In conclusion, we have discussed the feasibility of engineering both level repulsion and level attraction between two magnons coupled to a single-mode cavity, by manipulating the relaxation rates and the magnon-photon coupling strengths. In the weak-coupling regime, when the cavity mode relaxes considerably faster than the magnon modes, an effective dissipative interaction is engendered between the two magnons, which is directly responsible for the unfolding of level attraction as well as a narrow transmission window. The level crossing is manifested as the coalescence of polaritonic peaks in the spincurrent response of the hybridized system. On the contrary, the strong-coupling regime, for comparable relaxation rates, results in distinguishable polaritonic peaks with the earmarks of level repulsion. Interestingly, when the magnons are tuned in resonance to the cavity together with symmetrical couplings, the spectral response is characterized by the appearance of a dark polaritonic state. The central dip designating the dark polariton transitions into a peak as we deviate from resonance, indicating the transfer of population into the dark mode. It could be noted that the bulk of the precursive literature on level attraction in cavity-magnonics was grounded in the study of magnon-photon coupling and the resulting polariton characteristics. On the other hand, the level attraction highlighted in our paper, is rooted in an effective magnon-magnon coupling infused via the photonic interaction with the individual magnon modes. 
 
\section{Acknowledgements}
The authors acknowledge the support of The Air Force Office of Scientific Research [AFOSR award no FA9550-20-1-0366], The Robert A. Welch Foundation [grant no A-1943] and the Herman F. Heep and Minnie Belle Heep Texas A\&M University endowed fund.
\appendix
\section{Adiabatic Theory of Level Attraction}
\label{Appendix:a}

To explain the physics of our observation, in the limit $\kappa\gg\gamma_i$ and $\kappa\gg g_i$, we employ an adiabatic model to advance an effective two-mode description for the magnon modes, since the cavity mode relaxes rapidly into its steady state. This treatment reinforces forthwith the emergence of a dissipative interaction between the magnons in the adiabatic limit, which underpins the phenomenon of level attraction. In the rotating frame of the cavity mode, the mean field equations of the hybrid cavity magnon system are given by 
\begin{align}
\dot{X}=-i\mathcal{H}X,%+F_0e^{-i\Delta_a t},
\end{align} 
In the adiabatic limit, we can eliminate the cavity mode by setting $\dot{a}=0$, yielding 
\begin{align}
a=\frac{-i(g_1m_1+g_2m_2)}{\kappa}.
\end{align}
Substituting this into Eq. (A1), we obtain a coupled system of differential equations encompassing the dynamics of the magnon modes,
\begin{align}
\dot{Y}=-i\tilde{\mathcal{H}}Y,%+{F},
\end{align}
with ${Y}^{T}=$[$m_1$ $m_2$] and %, $F^{T}=$ [$\Omega$ $\Omega$] and 
\begin{align}
\tilde{\mathcal{H}}=\begin{pmatrix}
{s}-i\tilde{\gamma_1}&\frac{-ig_1g_2}{\kappa}\\
\frac{-ig_1g_2}{\kappa}&{-s}-i\tilde{\gamma_2}
\end{pmatrix},
\end{align}
where $\tilde{\gamma_i}=\gamma_i+\frac{g_i^2}{\kappa}$. For identical couplings $g_1=g_2=g$ and magnon relaxation rates $\gamma_1=\gamma_2=\gamma$, we can simplify $\tilde{\mathcal{H}}$ further into 
\begin{align}
\tilde{\mathcal{H}}=\begin{pmatrix}
s-i(\gamma+\frac{g^2}{\kappa})&\frac{-ig^2}{\kappa}\\
\frac{-ig^2}{\kappa}&-s-i(\gamma+\frac{g^2}{\kappa})
\end{pmatrix}.
\end{align}
The eigenvalues of the Hamiltonian are given by 
\begin{align}
\lambda_{\pm}=-i\bigg(\gamma+\frac{g^2}{\kappa}\bigg)\pm\sqrt{s^2-(g^2/\kappa)^2},
\end{align}
which bear the earmarks of level attraction, and also reproduce Eq. (6) in the limit of vanishingly small $\gamma$. For, $\abs{s}\leq \frac{g^2}{\kappa}$, the real parts remain identical, yielding the phase transition points $s=\pm  \frac{g^2}{\kappa}$. So it is no coincidence that we had observed the onset of mode coalescence at this exact same point in the preceding section. Thus, under the purview of the adiabatic approximation, in which the short-lived nature of the intracavity field mode is accounted for, the full tripartite system reduces to an effective two-mode subsystem with conformable properties in the linear response.
The above matrix exactly replicates the dynamics of a two-mode system with density matrix $\rho$, coupled through an interposing reservoir, and described by the master equation \cite{agarwal2012quantum} 
\begin{equation}
\frac{\dd \rho}{\dd t}=-\frac{i}{\hbar}[{H},\rho]+\gamma_{1}\mathcal{L}(m_1)\rho+\gamma_{2}\mathcal{L}(m_2)\rho+2\Gamma\mathcal{L}(c)\rho,
\end{equation}
with $H=\Delta_1m_1^{\dagger}m_1+\Delta_2m_2^{\dagger}m_2$, $\Gamma=g^2/\kappa$, $c=\frac{1}{\sqrt{2}}(m_1+m_2)$, and the Liouvillian $\mathcal{L}$ is defined by $\mathcal{L}(\sigma)\rho=2\sigma\rho \sigma^{\dagger}-\sigma^{\dagger}\sigma\rho-\rho \sigma^{\dagger}\sigma$. More precisely, the mean-field equations resulting from the above master equation are identical to Eq. (A3) with an effective Hamiltonian given by $\tilde{\mathcal{H}}$. To put it into perspective, the cavity, when weakly coupled to the magnons, merely acts as a common reservoir for the magnon modes in the adiabatic limit, thereby precipitating in a dissipative form of interaction between them.
\bibliography{references}

\end{document}